\documentclass[12pt]{article}
\usepackage{amsfonts}
\newtheorem{lemma}{Lemma}
\newtheorem{theorem}{Theorem}
\newtheorem{proposition}{Proposition}

\newenvironment{proof}{{\bf Proof}.}{\hfill $\Box$}

\begin{document}


\title{\sc Quadratic exponential vectors}

\author{Luigi Accardi, Ameur Dhahri\\
\vspace{-2mm}
\scriptsize Volterra Center, University of Roma Tor Vergata\\
\vspace{-2mm}
\scriptsize Via Columbia 2, 00133 Roma, Italy
\\\vspace{-2mm}\scriptsize e-mail:accardi@volterra.uniroma2.it\\
\scriptsize ameur@volterra.uniroma2.it}
\date{}
\maketitle
\begin{abstract}
We give a sufficient condition for the existence 
of a quadratic exponential vector with test function in 
$L^2(\mathbb{R}^d)\cap L^\infty(\mathbb{R}^d)$.
We prove the linear independence and totality, in the quadratic Fock space, 
of these vectors. 
Using a technique different from the one used in \cite {ADS},
we also extend, to a more general class of test functions, 
the explicit form of the scalar product between two such vectors.
\end{abstract}

\section{Introduction}\label{sec-intr}

Exponential vectors play a fundamental role in first order quantization.
It is therefore natural to expect that their quadratic analogues, first introduced
in \cite {AS}, will 
play such a role in the only example, up to now, of nonlinear renormalized 
quantization whose structure is explicitly known: the quadratic Fock functor.\\
The canonical nature of the objects involved justifies a detailed study of their
structure. \\
Let us emphasize that the nonlinearity introduces substantially new features
with respect to the linear case so that, contrarily to what happens in the usual
deformations of the commutation relations, quadratic quantization is not a simple
variant of the first order one, but interesting new phenomena arise.\\
For example usual exponential vectors can be defined for any
square integrable test function, but this is by far not true for quadratic ones.
This poses the problem to characterise those test functions for which
quadratic exponential vectors can be defined. This is the first problem
discussed in the present paper (see Theorem (\ref{exist-QEV})).\\
Another problem is the linear independence of the quadratic exponential vectors.
This very useful property, in the first order case, is a simple consequence 
of the linear independence of the complex exponential functions.
In the quadratic case the proof is more subtle. This is the second main result 
of the present paper (see Theorem (\ref{LI-quad-exp-v})).\\
For notations and terminology we refer to the papers
cited in the bibliography.
We simply recall that the algebra of the renormalized square of white noise (RSWN) 
with test function algebra
$$
{\cal A} := L^2(\Bbb R^d)\cap L^\infty(\Bbb R^d)
$$
is the $*$-Lie-algebra, with self--adjoint central element denoted $1$, generators 
$$
\{ B^+_f,B_h, N_g, 1 \ : \ f,g,h\in L^2(\Bbb R^d)\cap L^\infty(\Bbb R^d)\}
$$ 
involution 
$$
(B^+_f)^*=B_f \qquad , \qquad N_f^*=N_{\bar{f}}
$$ 
and commutation relations
$$
[B_f,B^+_g]=2c\langle f,g\rangle+4N_{\bar fg},\,\;[N_a,B^+_f]=2B^+_{af},\;c>0
$$
$$
[B^+_f,B^+_g]=[B_f,B_g]=[N_a,N_{a'}]=0
$$
for all $a$, $a'$, $f$, $g\in L^2(\Bbb R^d)\cap L^\infty(\Bbb R^d)$. 
Such algebra admits a unique, up to unitary isomorphism, $*$--representation
(in the sense defined in \cite{[AcKuSt09sigma]}) characterized by the existence of a
cyclic vector $\Phi$ satisfying
$$
B_h \Phi=  N_g\Phi = 0  \qquad : \qquad g,h\in L^2(\Bbb R^d)\cap L^\infty(\Bbb R^d)\}
$$ 
(see \cite{AcDha09a} for more detailed properties of this representation).

\section{Factorisation properties of the quadratic exponential vectors}\label{Fact-prop}

Recall from \cite {AS} that, if a quadratic exponential vector with test function
$f\in L^2(\mathbb{R}^d)\cap L^\infty(\mathbb{R}^d)$ exists, it is given by
$$
\Psi(f)=\sum_{n\geq0}\frac{B^{+n}_f\Phi}{n!}
$$ 
where by definition
\begin{eqnarray*}
\Psi(0):= B^{+0}_f\Phi := \Phi
\end{eqnarray*}
(notice the absence of the square root in the denominator).\\
In this section, after proving some simple consequences of the commutation 
relations, we give a direct proof for the factorization property of 
the exponential vectors.

\begin{lemma}\label{1} 
For all $f,g\in L^2(\Bbb R^d)\cap L^\infty(\Bbb R^d)$, one has
\begin{equation}\label{eq1} 
[N_f,B^{+n}_g]=2nB^{+(n-1)}_gB^+_{fg},
\end{equation}
\begin{equation}\label{eq2} 
[B_f,B^{+n}_g]=2nc\langle f,g\rangle B^{+(n-1)}_g+4nB^{+(n-1)}_gN_{\bar fg}+4n(n-1)B^{+(n-2)}_gB^+_{\bar fg^2}.
\end{equation}
\end{lemma}

\begin{proof} The mutual commutativity of the creators implies that

\begin{eqnarray*}
[N_f,B^{+n}_g]&=&\sum^{n-1}_{i=0}B^{+i}_g[N_f,B^+_g]B^{+(n-i-1)}_g\\
&=&\sum^{n-1}_{i=0}B^{+i}_g(2B^+_{fg})B^{+(n-i-1)}_g=2nB^{+(n-1)}_gB^+_{fg}
\end{eqnarray*}

which is (\ref{eq1}). To prove (\ref{eq2}) consider the identity
\begin{eqnarray*}
[B_f,B^{+n}_g] &=&\sum^{n-1}_{i=0}B^{+i}_g[B_f,B^+_g]B^{+(n-i-1)}_g \\
&=&\sum^{n-1}_{i=0}B^{+i}_g(2c\langle f,g\rangle+4N_{\bar fg})B^{+(n-i-1)}_g 
\\
&=& 2nc\langle f,g\rangle B^{+(n-1)}_g+4\sum^{n-1}_{i=0}B^{+i}_gN_{\bar fg}B^{+(n-i-1)}_g.
\end{eqnarray*}  
From (\ref{eq1}) it follows that 
$$
N_{\bar fg}B^{+(n-i-1)}_g=B^{+(n-i-1)}_gN_{\bar fg}+2(n-i-1)B^{+(n-i-2)}_gB^+_{\bar fg^2}.$$
Therefore, one obtains 
\begin{eqnarray*}
[B_f, B^{+n}_g] &=& 2nc\langle f,g\rangle B^{+(n-1)}_g+4nB^{+(n-1)}_gN_{\bar fg}\\
  &&+4\sum^{n-1}_{i=0}B^{+i}_g(2(n-i-1))B^{+(n-i-2)}_gB^+_{\bar fg^2}\\
 &= &2nc\langle f,g\rangle B^{+(n-1)}_g+4nB^{+(n-1)}_gN_{\bar fg}\\
  &&+8\left(\sum^{n-1}_{i=0}(n-i-1)\right)B^{+(n-2)}_gB^+_{\bar fg^2}\\
 &=& 2nc\langle f,g\rangle B^{+(n-1)}_g+4nB^{+(n-1)}_gN_{\bar fg}\\
 &&+8\left(\sum^{n-1}_{i=0}(n-i-1)\right)B^{+(n-2)}_gB^+_{\bar fg^2}\\
 &=& 2nc\langle f,g\rangle B^{+(n-1)}_g+4nB^{+(n-1)}_gN_{\bar fg}\\
  &&+8\left(n(n-1)-{n(n-1)\over2}\right)B^{+(n-2)}_gB^+_{\bar fg^2}\\
 &=& 2nc\langle f,g\rangle B^{+(n-1)}_g+4nB^{+(n-1)}_gN_{\bar fg}\\
 && +4n(n-1)B^{+(n-2)}_gB^+_{\bar fg^2}.
\end{eqnarray*}
\end{proof}   
\begin{lemma} \label{3} 
Let $f_1,\dots,f_k$, $g_1,\dots,g_h\in L^2(\Bbb R^d)\cap L^\infty(\Bbb R^d)$. 
Then, one has
$$B_{f_k}\dots B_{f_1}B^+_{g_h}\dots B^+_{g_1}\Phi=0\hbox{ for all }k>h\geq0.
$$
\end{lemma}

\begin{proof} It is sufficient to prove that
\begin{equation}
B_{f_{h+1}}\dots B_{f_1}B^+_{g_h}\dots B^+_{g_1}\Phi=0\ ,\qquad\forall\,h\in\Bbb N.\label{(1)}
\end{equation}
Suppose by induction that, for $h\geq0$, (\ref{(1)}) is satisfied. Then, one has
\begin{eqnarray*}
B_{f_{h+1}}\dots B_{f_1}B^+_{g_h}\dots B^+_{g_1}\Phi\!\!&=&\!\!
B_{f_{h+1}}\dots B_{f_2}[B_{f_1},B^+_{g_h}\dots B^+_{g_1}]\Phi\\
&=&\!\!\!\!B_{f_{h+1}}\dots B_{f_2}\sum^1_{i=h}B^+_{g_h}\dots B^+_{g_{i+1}}(2c\langle f_1,g_i\rangle+4N_{\bar f_1g_i})\\
&&\;\;\;\;\;\;\;\;\;\;\;\;B^+_{g_{i-1}}\dots B^+_{g_1}\Phi\\
&=&\!\!2c\sum^1_{i=h}\langle f_1,g_i\rangle(B_{f_{h+1}}\dots B_{f_2})(B^+_{g_h}\dots\hat B^+_{g_i}\dots B^+_{g_1})\Phi\\
&&+4\sum^1_{i=h}(B_{f_{h+1}}\dots B_{f_2})(B^+_{g_h}\dots 
N_{\bar f_1g_i}B^+_{g_{i-1}}\dots B^+_{g_1})\Phi.
\end{eqnarray*} 
By the induction assumption
$$
\sum^1_{i=h}\langle f_1,g_i\rangle(B_{f_{h+1}}\dots B_{f_2})(B^+_{g_h}\dots\hat B^+_{g_i}\dots B^+_{g_1})\Phi=0.$$
Therefore, one gets 
\begin{eqnarray*}
& &B_{f_{h+1}}\dots B_{f_1}B^+_{g_h}\dots B^+_{g_1}\Phi\\
&&=4\sum^1_{i=h}(B_{f_{h+1}}\dots B_{f_2})(B^+_{g_h}\dots B^+_{g_{i+1}}N_{\bar f_1g_i}B^+_{g_{i-1}}\dots B^+_{g_1})\Phi\\
&&=4\sum^1_{i=h}(B_{f_{h+1}}\dots B_{f_2})\left(B^+_{g_h}\dots B^+_{g_{i+1}}\sum^1_{m=i-1}B^+_{g_{i-1}}\dots[N_{\bar f_1g_i},B^+_{g_m}]\dots B^+_{g_1}\right)\Phi\\
&&=4\sum^1_{i=h}B_{f_{h+1}}\dots B_{f_2}B^+_{g_h}\dots B^+_{g_{i+1}}\left(2\sum^1_{m=i-1}B^+_{g_{i-1}}\dots B^+_{\bar f_1g_ig_m}B^+_{g_{m-1}}\dots B^+_{g_1}\right)\Phi\\
&&=8\sum^1_{i=h}\sum^1_{m=i-1}B_{g_{h+1}}\dots B_{f_2}B^+_{g_h}\dots B^+_{g_{i+1}}\dots \hat{B}^+_{g_{m}}\dots B^+_{g_1}B^+_{\bar f_1g_ig_m}\Phi,
\end{eqnarray*}
which is equal to $0$ by the induction assumption. 
\end{proof}

\begin{lemma} \label{4}
Let $I$, $J\subset\Bbb R^d$ such that $I\cap J=\phi$ and let $f_1,\dots,f_h$, $f'_1,\dots,f'_{h'}$, $g_1,\dots,g_k$, $g'_1,\dots,g'_{k'}\in L^2(\Bbb R^d)\cap L^\infty(\Bbb R^d)$ such that 
$$
\hbox{supp }(f_i)\subset I \ , \ \hbox{supp }(f'_i)\subset I \ , \
\hbox{supp }(g_i)\subset J \ , \ \hbox{supp }(g'_i)\subset J
$$
 Then, for $h\not=h'$ or $k\not=k'$, one has
$$
\langle B^+_{f_h}\dots B^+_{f_1}B^+_{g_k}\dots B^+_{g_1}\Phi,B^+_{f'_{h'}}\dots B^+_{f'_1}
B^+_{g'_{k'}}\dots B^+_{g'_1}\Phi\rangle=0.
$$
\end{lemma}

\begin{proof} 
Lemma (\ref{1})  and the polarization identity imply that \\
$[B_{f_h}\dots B_{f_1}$, $B^+_{g'_{k'}}\dots B^+_{g'_1}]=0$. 
Therefore, it is sufficient to prove the result for $h\not=h'$. 
Taking eventually the complex conjugate, 
we can suppose that $h>h'$. Under this assumption
\begin{eqnarray*}
&&\langle B^+_{f_h}\dots B^+_{f_1}B^+_{g_k}\dots B^+_{g_1}\Phi,B^+_{f'_{h'}}\dots B^+_{f'_1}B^+_{g'_{k'}}\dots B^+_{g'_1}\Phi\rangle\\
&&=\langle B^+_{g_k}\dots B^+_{g_1}\Phi,B^+_{g'_{k'}}\dots B^+_{g'_1}(B_{f_h}\dots B_{f_1}B^+_{f'_{h'}}\dots B^+_{f'_1})\Phi\rangle
\end{eqnarray*}
and the statement follows because, from Lemma \ref{3}, one has
$$
B_{f_h}\dots B_{f_1}B^+_{f'_{h'}}\dots B^+_{f'_1}\Phi=0.
$$
\end{proof}

For $I\subset\Bbb R^d$, denote ${\cal H}_I$ the closed linear span of the set\\ $\big\{B^{+n}_f\Phi$, $n\in\Bbb N$, $f\in L^2(\Bbb R^d)\cap L^\infty(\Bbb R^d)$ such that $\hbox{supp }(f)\subset I\big\}$ \\
and ${\cal H}={\cal H}_{\Bbb R^d}$. The space $\mathcal{H}$, denoted 
$\Gamma_2(L^2(\mathbb{R}^d)\cap L^\infty(\mathbb{R}^d))$, is called the 
{\it quadratic Fock space} with test function algebra 
$L^2(\mathbb{R}^d)\cap L^\infty(\mathbb{R}^d)$.
\begin{lemma} \label{5}
Let $I,J\subset\Bbb R^d$ such that $I\cap J=\phi$ and let $f_1,\dots,f_h$, $f'_1,\dots,f'_h$, $g_1,\dots,g_k$, $g'_1,\dots,g'_k\in L^2(\Bbb R^d)\cap L^\infty(\Bbb R^d)$ such that\\
 $\hbox{supp }(f_i)\subset I$, $\hbox{supp }(f'_i)\subset I$, $\hbox{supp }(g_i)\subset J$ and $\hbox{supp }(g'_i)\subset J$. \\
 Then, one has
\begin{eqnarray}\label{(2)}
&&\langle B^+_{f_h}\dots B^+_{f_1}B^+_{g_k}\dots B^+_{g_1}\Phi,B^+_{f'_h}\dots B^+_{f'_1}B^+_{g'_k}\dots B^+_{g'_1}\Phi\rangle\nonumber\\
&&=\langle B^+_{f_h}\dots B^+_{f_1}\Phi_I,B^+_{f'_h}\dots B^+_{f'_1}\Phi_I\rangle
\langle B^+_{g_k}\dots B^+_{g_1}\Phi_J,B^+_{g'_k}\dots B^+_{g'_1}\Phi_J\rangle.
\end{eqnarray}
\end{lemma}
\begin{proof} By induction on $h$. For $h=1$, one has
\begin{eqnarray*}
&&\langle B^+_{f_1}B^+_{g_k}\dots B^+_{g_1}\Phi,B^+_{f'_1}B^+_{g'_k}\dots B^+_{g'_1}\Phi\rangle\\
&&=\langle B^+_{g_k}\dots B^+_{g_1}\Phi,B^+_{g'_k}\dots B^+_{g'_1}(B_{f_1}B^+_{f'_1})\Phi\rangle \\
&&=\!2c\langle f_1,f'_1\rangle\langle B^+_{g_k}\dots B^+_{g_1}\Phi,B^+_{g'_k}\dots B^+_{g'_1}\Phi\rangle\\
&&=\langle B^+_{f_1}\Phi_I,B^+_{f'_1}\Phi_I\rangle\langle B^+_{g_k}\dots B^+_{g_1}\Phi_J,B^+_{g'_k}\dots B^+_{g'_1}\Phi_J\rangle.
\end{eqnarray*}
Let $h\geq1$ and suppose that (\ref{(2)}) holds true. Then, one has
\begin{eqnarray*}
&&\langle B^+_{f_{h+1}}B^+_{f_h}\dots B^+_{f_1}B^+_{g_k}\dots B^+_{g_1}\Phi,B^+_{f'_{h+1}}B^+_{f'_h}\dots B^+_{f'_1}B^+_{g'_k}\dots B^+_{g'_1}\Phi\rangle\\
&&=\langle B^+_{f_h}\dots B^+_{f_1}B^+_{g_k}\dots B^+_{g_1}\Phi,B_{f_{h+1}}B^+_{f'_{h+1}}B^+_{f'_h}\dots B^+_{f'_1}B^+_{f'_1}B^+_{g'_k}\dots B^+_{g'_1}\Phi\rangle\\
\!\!&&\!\!=\!\!\sum^1_{m=h+1}\langle B^+_{f_h}\dots B^+_{f_1}B^+_{g_k}\dots B^+_{g_1}\Phi,(B^+_{g'_k}\dots B^+_{g'_1})(B^+_{f'_{h+1}}\dots[B_{f_{h+1}},B^+_{f'_m}]\dots B^+_{f'_1})\Phi\rangle\\
&&=2c\sum^1_{m=h+1}\langle f_{h+1},f'_m\rangle\\
&&\langle B^+_{f_h}\dots B^+_{f_1}B^+_{g_k}\dots B^+_{g_1}\Phi,(B^+_{g'_k}\dots B^+_{g'_1})(B^+_{f'_{h+1}}\dots\hat B^+_{f'_m}\dots B^+_{f'_1})\Phi\rangle\\
&&+4\sum^1_{m=h+1}\langle B^+_{f_h}\dots B^+_{f_1}B^+_{g_k}\dots B^+_{g_1}\Phi,\\
&&\;\;\;\;\;\;(B^+_{g'_k}\dots B^+_{g'_1})(B^+_{f'_{h+1}}\dots B^+_{f'_{m+1}}N_{\bar f_{h+1}f'_m}B^+_{f'_{m-1}}\dots B^+_{f'_1})\Phi\rangle.
\end{eqnarray*} 
By the induction assumption
\begin{eqnarray*}
&&\sum^1_{m=h+1}\langle f_{h+1},f'_m\rangle\langle B^+_{f_h}\dots B^+_{f_1}B^+_{g_k}\dots B^+_{g_1} \Phi,B^+_{g'_k}\dots B^+_{g'_1}B^+_{f'_{h+1}}\dots \hat B^+_{f'_m}\dots B^+_{f'_1}\Phi\rangle\\
&&=\sum^1_{m=h+1}\langle f_{h+1},f'_m\rangle\langle B^+_{f_h}\dots B^+_{f_1}\Phi_I,B^+_{f'_{h+1}}\dots \hat B^+_{f'_m}\dots B^+_{f'_1}\Phi_I\rangle\\
&&\;\;\;\;\;\;\langle B^+_{g_k}\dots B^+_{g_1} \Phi_J,B^+_{g'_k}\dots B^+_{g'_1}\Phi_J\rangle.
\end{eqnarray*}
Moreover
\begin{eqnarray*}
&&\langle B^+_{f_h}\dots B^+_{f_1}B^+_{g_k}\dots B^+_{g_1}\Phi,B^+_{g'_k}\dots B^+_{g'_1}B^+_{f'_{h+1}}\dots B^+_{f'_{m+1}}N_{\bar f_{h+1}f'_m}B^+_{f'_{m-1}}\dots B^+_{f'_1}\Phi\rangle\\
&&=\sum^1_{i=m-1}\langle B^+_{f_h}\dots B^+_{f_1}B^+_{g_k}\dots B^+_{g_1}\Phi,\\
&&\;\;\;\;\;\;\;\;\;\;B^+_{g'_k}\dots B^+_{g'_1}B^+_{f'_{h+1}}\dots\hat B^+_{f'_m}\dots[N_{\bar f_{h+1}f'_{m'}},B^+_{f'_i}]\dots B^+_{f_1'}\Phi\rangle\\
&&=2\sum^1_{i=m-1}\langle B^+_{f_h}\dots B^+_{f_1}B^+_{g_k}\dots B^+_{g_1}\Phi,\\
&&\;\;\;\;\;\;\;\;\;\;B^+_{g'_k}\dots B^+_{g'_1}B^+_{g'_1}B^+_{f'_{h+1}}\dots\hat B^+_{f'_m}\dots\hat B^+_{f'_i}\dots B^+_{f_1}B^+_{\bar f_{h+1}f'_mf'_i}\Phi\rangle.
\end{eqnarray*}
Again by the induction assumption, this term is equal to
\begin{eqnarray*}
&&2\sum^1_{i=m-1}\langle B^+_{f_h}\dots B^+_{f_1}\Phi_I,B^+_{f'_{h+1}}\dots\hat B^+_{f'_m}\dots\hat B^+_{f'_i}\dots B^+_{f'_1}B^+_{\bar f_{h+1}f'_mf'_i}\Phi_I\rangle\\
&&\;\;\;\;\;\;\;\;\;\;\langle B^+_{g_k}\dots B^+_{g_1}\Phi_J,B^+_{g'_k}\dots B^+_{g'_1}\Phi_J\rangle.\\
\end{eqnarray*}
Hence, one obtains
\begin{eqnarray*}
&&\langle B^+_{f_{h+1}}B^+_{f_h}\dots B^+_{f_1}B^+_{g_k}\dots B^+_{g_1}\Phi,B^+_{f'_{h+1}}B^+_{f'_h}\dots B^+_{f'_1}B^+_{g'_k}\dots B^+_{g'_1}\Phi\rangle\\
&&=\biggl\{\sum^1_{m=h+1}\Big[2c\langle f_{h+1},f'_m\rangle\langle B^+_{f_h}\dots B^+_{f_1}\Phi_I,B^+_{f'_{h+1}}\dots\hat B^+_{f'_m}\dots B^+_{f'_1}\Phi_I\rangle\\
&&+8\sum^1_{i=m-1}\langle B^+_{f_h}\dots B^+_{f_1}\Phi_I,B^+_{f'_{h+1}}\dots\hat B^+_{f'_m}\dots\hat B^+_{f'_i}\dots B^+_{f'_1}B^+_{\bar f_{h+1}f'_mf'_i}\Phi_I\rangle\Big]\biggl\}\\
&&\;\;\;\;\langle B^+_{g_k}\dots B^+_{g_1}\Phi_J,B^+_{g'_k}\dots B^+_{g'_1}\Phi_J\rangle.
\end{eqnarray*}
Since the term between square brackets is equal to
$$
\langle B^+_{f_{h+1}}\dots B^+_{f_1}\Phi_I,B^+_{f'_{h+1}}\dots B^+_{f'_1}\Phi_I\rangle.
$$
This completes the proof of the lemma. 
\end{proof}
\begin{lemma}\label{6} Let $I$, $J\subset\Bbb R^d$ such that $I\cap J=\phi$. 
Then, the operator
$$
U_{I,J}:{\cal H}_{I\cup J}\to{\cal H}_{I}\otimes{\cal H}_J
$$
defined by 
$$
U_{I,J}\big(B^+_{f_n}\dots B^+_{f_1}B^+_{g_m}\dots B^+_{g_1}\Phi_{I\cup J}\big)=(B^+_{f_n}\dots B^+_{f_1}\Phi_I)\otimes(B^+_{g_m}\dots B^+_{g_1}\Phi_J) 
$$
where $\hbox{supp }(f_i)\subset I$, $\hbox{supp }(g_i)\subset J$, is unitary.
\end{lemma}

\begin{proof} It is sufficient to prove that for all $f_n,\dots,f_1,f'_k,\dots,f'_1\in{\cal H}_I;$\\
$\ g_m,\dots,g_1,g'_h,\dots,g'_1\in{\cal H}_{J}$
\begin{eqnarray}\label{(3)}
&&\langle B^+_{f_n}\dots B^+_{f_1}B^+_{g_m}\dots B^+_{g_1}\Phi_{I\cup J},B^+_{f'_k}\dots B^+_{f'_1}B^+_{g'_h}\dots B^+_{g'_1}\Phi_{I\cup J}\rangle\nonumber\\
&&=\langle B^+_{f_n}\dots B^+_{f_1}\Phi_I,B^+_{f'_k}\dots B^+_{f'_1}\Phi_I\rangle\langle B^+_{g_m}\dots B^+_{g_1}\Phi_J,B^+_{g'_h}\dots B^+_{g'_1}\Phi_J\rangle.
\end{eqnarray}
Note that Lemma \ref{4} implies that (\ref{(3)}) holds if $n\not=h$ or $m\not=h$. Moreover, if $n=k$ and $m=h$, then from Lemma \ref{5}, the identity (\ref{(3)}) is also satisfied. This ends the proof. 
\end{proof}
\begin{theorem} \label{theo1}
Let $I_1,\dots,I_n\subset\Bbb R^d$ such that $I_i\cap I_j=\phi$ for all $i\not=j$. Let 
$$U_{I_1\cup\dots\cup I_{k-1},I_k}:{\cal H}_{\bigcup^k_{i=1}I_i}\to{\cal H}_{\bigcup^{k-1}_{i=1}I_i}\otimes{\cal H}_{I_k}$$
be the operator defined by Lemma \ref{6}. Then, the operator
$$U_{I_1,\dots,I_n}:{\cal H}_{\bigcup^n_{i=1}I_i}\to{\cal H}_{I_1}\otimes\dots\otimes{\cal H}_{I_n}$$
given by
$$U_{I_1,\dots,I_n}=(U_{I_1,I_2}\otimes1_{I_3}\otimes\dots\otimes1_{I_n})\circ(U_{I_1\cup I_2,I_3}\otimes1_{I_4}\otimes\dots\otimes1_{I_n})\circ\dots
$$
$$
\dots\circ (U_{I_1\cup\dots\cup I_{n-2},I_{n-1}}\otimes1_{I_n})\circ 
U_{I_1\cup\dots\cup I_{n-1},I_n}
$$
is unitary.
\end{theorem}
\begin{proof} For all $k\in\{2,\dots,n\}$
$(I_1\cup\dots\cup I_{k-1})\cap I_k=\emptyset$. Therefore the operator
$$
U_{I_1\cup\dots\cup I_{k-1},I_k}:{\cal H}_{\bigcup^k_{i=1}I_i}
\to{\cal H}_{\bigcup^{k-1}_{i=1}I_i}\otimes{\cal H}_{I_k}
$$
is unitary because of Lemma \ref{6}.
In particular $U_{I_1,\dots,I_n}$ is a unitary operator. 
\end{proof}
\begin{theorem} \label{theo2}
(factorization property of the quadratic exponential vectors)
Let $I_1,\dots,I_n\subset\Bbb R^d$ such that $I_i\cap I_j=\phi$, for all $i\not=j$ and let\\
$U_{I_1,\dots,I_n}:{\cal H}_{\bigcup^n_{i=1}I_i}\to
\bigotimes_{i=1}^n{\cal H}_{I_i}\otimes\dots\otimes{\cal H}_{I_n}$ 
be the unitary operator defined by Theorem \ref{theo1}.\\ 
Then, for all
$
f\in L^2\left(\bigcup^n_{i=1}I_i\right)\cap L^\infty\left(\bigcup^n_{i=1}I_i\right)
$
such that $\Psi(f)$ exists, one has
$$
U_{I_1,\dots,I_n}\Psi(f)=\Psi(f_{I_1})\otimes\dots\otimes\Psi(f_{I_n}) 
$$
where
$
f_{I_i}:=f\chi_{I_i}
$.
\end{theorem}
\begin{proof} Denote $J_k=\bigcup^k_{i=1}I_i$. Then, 
$U_{I_1\cup\dots\cup I_{k-1},I_k}=U_{J_{k-1},I_k}$ is a unitary 
operator from ${\cal H}_{J_k}$ to ${\cal H}_{J_{k-1}}\otimes{\cal H}_{I_k}.$ Let 
$
f\in L^2\left(\bigcup^n_{i=1}I_i\right)\cap L^\infty\left(\bigcup^n_{i=1}I_i\right)
$
be such that $\Psi(f)$ exists. Since
$$
f=\sum^n_{i=1}f_{I_i}=f_{J_{n-1}}+f_{I_n}
$$
it follows that
$$
B^{+m}_f=(B^+_{f_{J_{n-1}}}+B^+_{f_{I_n}})^m
=\sum^m_{k=0}C^k_mB^{+k}_{f_{J_{n-1}}}B^{+(m-k)}_{f_{I_n}} 
$$
where 
$$
C_m^k=\frac{m!}{(m-k)!k!}.
$$
Therefore, one obtains
$$
U_{J_{n-1},I_n}(B^+_{f_{J_{n-1}}}+B^+_{f_{I_n}})^m\Phi
=\sum^m_{k=0}C^k_m(B^{+k}_{f_{J_{n-1}}}\Phi_{J_{n-1}})\otimes B^{+(n-k)}_{f_{I_n}}\Phi_{I_n}.
$$
This gives 
$$
U_{J_{n-1},I_n}\Psi(f)=\Psi(f_{J_{n-1}})\otimes\Psi(f_{I_n}).
$$
Now, one has
$$
(U_{J_{n-2},I_{n-1}}\otimes1_{I_n})\circ(U_{J_{n-1},I_n})\Psi(f)
=(U_{J_{n-2},I_{n-1}}\Psi(f_{J_{n-1}}))\otimes\Psi(f_{I_n}).
$$
In the same way, we prove that
$$
U_{J_{n-2},I_{n-1}}\Psi(f_{J_{n-1}})=\Psi(f_{J_{n-2}})\otimes\Psi(f_{I_{n-1}}).
$$
Hence, the following identity holds
$$
(U_{J_{n-2},I_{n-1}}\otimes1_{I_n})\circ(U_{J_{n-1},I_n})\Psi(f)=\Psi(f_{J_{n-2}})\otimes\Psi(f_{I_{n-1}})\otimes\Psi(f_{I_n}).
$$
Iterating this procedure one finds
$$
(U_{J_k,I_{k+1}}\otimes1_{I_{k+2}}\otimes\dots\otimes1_{I_n})
\circ(U_{J_{k+1},I_{k+2}}\otimes 1_{I_{k+3}}\otimes\dots\otimes1_{I_n})
$$
$$
\circ\dots\circ U_{J_{n-1},I_n}\Psi(f)=\Psi(f_{J_k})\otimes\Psi(f_{I_{k+1}})
\otimes\dots\otimes\Psi(f_{I_n})
$$
or equivalently
$$
U_{I_1,\dots,I_n}\Psi(f)=(U_{I_1,I_2}\otimes1_{I_3}\otimes\dots\otimes 1_{I_n})\circ(U_{J_2,I_3}
\otimes1_{I_4}\otimes\dots\otimes1_{I_n})
$$
$$
\circ\dots\circ U_{j_{n-1},I_n}\Psi(f)=\Psi(f_{I_1})\otimes\dots\otimes\Psi(f_{I_n})
.$$
\end{proof}

\section{Condition for the existence of the  quadratic exponential vectors}
\label{exist-q-exp-vect}

When both $\Psi(f)$ and $\Psi(g)$ exist, the explicit form of their scalar product was determined in \cite{AS}, for step functions in $\mathbb{R}$ with bounded supports. We further give a sufficient condition for the existence of a quadratic exponential vector. Finally, using the factorization property of the quadratic exponential vectors and an approximation argument, we extend the formula for the scalar product to exponential vectors with arbitrary step functions. Due to the nonlinearity involved in this form of the scalar product, the approximation argument is not as straightforward as in the first order case. A different proof of this result was obtained in \cite{ADS}.

\begin{lemma}\label{hfg}
For all $n\geq 1$ and all $f,\,g,\,h\in L^2(\Bbb R^d)\cap L^\infty(\Bbb R^d)$, one has
\begin{eqnarray}\label{hhh}
\langle B^{+(n-1)}_f\Phi,B_hB^{+n}_g\Phi\rangle&=&c\sum^{n-1}_{k=0}2^{2k+1}{n!(n-1)!\over((n-k-1)!)^2}\,\langle hf^{k}, g^{k+1}\rangle\nonumber\\
\;\;\;\;\;&&\langle B^{+(n-k-1)}_f\Phi,B^{+(n-k-1)}_g\Phi\rangle.
\end{eqnarray}
\end{lemma}
\begin{proof} Let us prove the above lemma by induction. For $n=1$, it is clear that identity (\ref{hhh}) is satisfied. 

Now, let $n\geq1$ and suppose that (\ref{hhh}) holds true. Note that, from (\ref{eq2}) it follows that
$$
B_hB^{+(n+1)}_g\Phi=2(n+1)c\langle h,g\rangle B^{+n}_g\Phi+4n(n+1)B^+_{\bar{h}g^2}B^{+(n-1)}_g\Phi.
$$
Then, one gets
\begin{eqnarray}\label{over}
\langle B^{+n}_f\Phi,B_hB^{+(n+1)}_g\Phi\rangle&=&2(n+1)c\langle h,g\rangle\langle B^{+n}_f\Phi,B^{+n}_g\Phi\rangle\nonumber\\
&&+4n(n+1)\langle B^{+n)}_f\Phi,B^+_{\bar{h}g^2}B^{+(n-1)}_g\Phi\rangle\nonumber\\
&=&2(n+1)c\langle f,g\rangle\langle B^{+n}_f\Phi,B^{+n}_g\Phi\rangle\nonumber\\
&&+4n(n+1)\overline{\langle B^{+(n-1)}_g\Phi, B_{\bar{h}g^2}B^{+n}_f\Phi\rangle}.
\end{eqnarray}
Therefore, by induction assumption, one has
\begin{eqnarray*}
\langle B^{+(n-1)}_g\Phi, B_{\bar{h}g^2}B^{+n}_f\Phi\rangle&=&c\sum_{k=0}^{n-1}2^{2k+1}\frac{n!(n-1)!}{((n-k-1)!)^2}\langle (\bar{h}g^2)g^k,f^{k+1}\rangle\\
\;\;\;\;\;&&\langle B^{+(n-k-1)}_g\Phi,B^{+(n-k-1)}_f\Phi\rangle\\
&=&c\sum_{k=1}^n2^{2k-1}\frac{n!(n-1)!}{((n-k)!)^2}\langle g^{k+1},hf^k\rangle\\
\;\;\;\;\;&&\langle B^{+(n-k)}_g\Phi,B^{+(n-k)}_f\Phi\rangle.
\end{eqnarray*}
It follows that
\begin{eqnarray}\label{aaa}
4n(n+1)\overline{\langle B^{+(n-1)}_g\Phi, B_{\bar{h}g^2}B^{+n}_f\Phi\rangle}&=&c\sum_{k=1}^n2^{2k+1}\frac{n!(n+1)!}{((n-k)!)^2}\,\langle hf^{k}, g^{k+1}\rangle\nonumber\\
\;\;\;\;\;&&\langle B^{+(n-k)}_f\Phi,B^{+(n-k)}_g\Phi\rangle.
\end{eqnarray}  
Finally, identities (\ref{over}) and (\ref{aaa}) imply that
\begin{eqnarray*}
\langle B^{+n}_f\Phi,B_hB^{+(n+1)}_g\Phi\rangle&=&c\sum^{n}_{k=0}2^{2k+1}{n!(n+1)!\over((n-k)!)^2}\,\langle hf^{k}, g^{k+1}\rangle\\
\;\;\;\;\;&&\langle B^{+(n-k)}_f\Phi,B^{+(n-k)}_g\Phi\rangle.
\end{eqnarray*}
This ends the proof.
\end{proof}

From the above lemma, we prove the following.
\begin{proposition} \label{prop1}
For all $n\geq 1$ and all $f,\,g\in L^2(\Bbb R^d)\cap L^\infty(\Bbb R^d)$, one has
\begin{eqnarray*}
\langle B^{+n}_f\Phi,B^{+n}_g\Phi\rangle&=&c\sum^{n-1}_{k=0}2^{2k+1}{n!(n-1)!\over((n-k-1)!)^2}\,\langle f^{k+1}, g^{k+1}\rangle\\
\;\;\;\;\;&&\langle B^{+(n-k-1)}_f\Phi,B^{+(n-k-1)}_g\Phi\rangle.
\end{eqnarray*}
\end{proposition}
\begin{proof} 
In order to prove the above proposition, it is sufficient to take $f=h$ in Lemma \ref{hfg}.
\end{proof}

As a consequence of the Proposition \ref{prop1}, we give a sufficient condition for the existence of an exponential vector with a given test function.
\begin{theorem}\label{exist-QEV}
Let $f\in L^2(\mathbb{R}^d)\cap L^2(\mathbb{R}^d)$. The quadratic exponential vector $\Psi(f)$ exists if $\|f\|_\infty<\frac{1}{2}$, and does not exists if  $\|f\|_\infty>\frac{1}{2}$.
\end{theorem}
\begin{proof}
Sufficiency. Let $f\in L^2(\mathbb{R}^d)\cap L^2(\mathbb{R}^d)$ be such that $||f||_\infty<\frac{1}{2}$. From the above proposition, one has
\begin{eqnarray}\label{tun}
||B^{+n}_f\Phi||^2&=&c\sum_{k=0}^{n-1}2^{2k+1}\frac{n!(n-1)!}{((n-k-1)!)^2}|\|f^{k+1}\|^2_2\|B_f^{+(n-k-1)}\Phi\|^2\nonumber\\
&=&c\sum_{k=1}^{n-1}2^{2k+1}\frac{n!(n-1)!}{((n-k-1)!)^2}|\|f^{k+1}\|^2_2\|B_f^{+(n-k-1)}\Phi\|^2\nonumber\\
&&+2nc\|f\|^2_2\|B^{+(n-1)}_f\Phi\|^2  +2nc\|f\|^2_2\|B^{+(n-1)}_f\Phi\|^2 \nonumber\\
 &=& c\sum_{k=0}^{n-2}2^{2k+3}\frac{n!(n-1)!}{(((n-1)-k-1)!)^2}|\|f^{k+2}\|^2_2\|B_f^{+((n-1)-k-1)}\Phi\|^2\nonumber\\
 &&\leq
\big(4n(n-1)\|f\|^2_\infty\big)\Big[c\sum_{k=0}^{n-2}2^{2k+1}
\frac{(n-1)!(n-2)!}{(((n-1)-k-1)!)^2}\nonumber\\
&&\;\;\;\;\;\;\;\;\;\;\;\;\;\;\;\|f^{k+1}\|^2_2
  B_f^{+((n-1)-k-1)}\Phi\|^2\Big].
\end{eqnarray}
Note that
$$||B^{+(n-1)}_f\Phi||^2=c\sum_{k=0}^{n-2}2^{2k+1}\frac{(n-1)!(n-2)!}{((n-1)-k-1)!)^2}|\|f^{k+1}\|^2_2\|B_f^{+((n-1)-k-1)}\Phi\|^2.$$
This proves that
\begin{eqnarray}\label{you}
||B^{+n}_f\Phi||^2\leq\Big[4n(n-1)\|f\|^2_\infty+2n\|f\|^2_2\Big]\|B^{+(n-1)}_f\Phi\|^2.
\end{eqnarray}
Finally, one gets
$$\frac{||B^{+n}_f\Phi||^2}{(n!)^2}\leq\Big[\frac{4n(n-1)\|f\|^2_\infty+2n\|f\|^2_2}{n^2}\Big]\frac{\|B^{+(n-1)}_f\Phi\|^2}{((n-1)!)^2}.$$
Hence, it is clear that if $4\|f\|_\infty^2<1$, then the series $\sum_{n\geq0}\frac{||B^{+n}_f\Phi||^2}{(n!)^2}$ converges.

Now, let $f\in L^2(\mathbb{R}^d)\cap L^\infty(\mathbb{R}^d)$ be such that $||f||_\infty>\frac{1}{2}$. Put
$$J=\{x\in\mathbb{R}^d,\;|f(x)|\geq \frac{1}{2}\}.$$
It is clear that $|J|>0$. In fact, if $|J|=0$, this implies that a.e $x\in\mathbb{R}^d$, $|f(x)|<\frac{1}{2}$ and $\|f\|_\infty\leq\frac{1}{2}$ , against our hypothesis. It follows that
\begin{eqnarray}\label{amr}
|f(x)|\geq \chi_J(x)|f(x)|\geq \frac{1}{2}\chi_J(x),
\end{eqnarray}
for almost all $x\in\mathbb{R}^d$. Note that from Proposition \ref{prop1}, it is easy to show by induction that if 
$$|h(x)|\geq |g(x)|,\;a.e$$
then 
\begin{equation}\label{ABD}
\|B^{+n}_h\Phi\|^2\geq\|B^{+n}_g\Phi\|^2.
\end{equation}
Hence, identities (\ref{amr}) and (\ref{ABD}) imply that
$$\sum_{n\geq0}\frac{1}{(n!)^2}\|B^{+n}_f\Phi\|^2\geq\sum_{n\geq0}\frac{1}{(n!)^2}\|B^{+n}_{\frac{1}{2}\chi_J}\Phi\|^2.$$
But, from Proposition \ref{prop1}, one has
$$
\frac{\|B^{+n}_{\frac{1}{2}\chi_J}\Phi\|^2}{(n!)^2} = \Big(\frac{c|J|}{2n}+\frac{n-1}{n}\Big)\frac{\|B^{+(n-1)}_{\frac{1}{2}\chi_J}\Phi\|^2}{((n-1)!)^2}\\
 \geq \frac{n-1}{n}\frac{\|B^{+(n-1)}_{\frac{1}{2}\chi_J}\Phi\|^2}{((n-1)!)^2}\geq \dots
$$
$$
\dots \geq \frac{n-1}{n}\frac{n-2}{n-1}\dots\frac{1}{2}\|B^{+}_{\frac{1}{2}\chi_J}\Phi\|^2=\frac{1}{n}\|B^{+}_{\frac{1}{2}\chi_J}\Phi\|^2=\frac{1}{4n}|J|^2,
$$  
which proves that the series $\sum_{n\geq0}\frac{1}{(n!)^2}\|B^{+n}_{\frac{1}{2}\chi_j}\Phi\|^2$ diverges. It follows that also the series $\sum_{n\geq0}\frac{1}{(n!)^2}\|B^{+n}_f\Phi\|^2$ diverges.
\end{proof}

Theorem \ref{exist-QEV} implies that, 
whenever the quadratic exponential vectors are well defined, their scalar product 
$$
\langle \Psi(f),\Psi(g)\rangle
$$
exists. More precisely the following results hold.
\begin{theorem}\label{dha}
For all $f,\,g\in L^2(\mathbb{R}^d)\cap L^\infty(\mathbb{R}^d)$ such that\\ $\|f\|_\infty<\frac{1}{2}$ and $\|g\|_\infty<\frac{1}{2}$, then the integral
$$
\int_{\mathbb{R}^d}\ln(1-4\bar{f}(s)g(s))ds
$$
exists. Moreover, one has
\begin{eqnarray}\label{identity}
\langle \Psi(f),\Psi(g)\rangle=e^{-\frac{c}{2}\int_{\mathbb{R}^d}\ln(1-4\bar{f}(s)g(s))ds}.
\end{eqnarray}
\end{theorem}
\begin{proof}
Proposition \ref{prop1} implies that, for any pair of step functions $f=\rho\chi_I$, $g=\sigma\chi_I,$ where $|\rho|<\frac{1}{2},\;|\sigma|<\frac{1}{2}$ 
and $I\subset\mathbb{R}^d$ such that its Lebesgue measure $|I|<\infty$, one has
\begin{eqnarray*}
\langle B^{+n}_f\Phi,B^{+n}_g\Phi\rangle&=&c\sum_{k=0}^{n-2}2^{2k+3}\frac{n!(n-1)!}{(((n-1)-k-1)!)^2}\langle f^{k+2},g^{k+2}\rangle\\
&&\langle B_f^{+((n-1)-k-1)}\Phi,B_g^{+((n-1)-k-1)}\Phi\rangle\\
&&+2nc\langle f,g\rangle\langle B^{+(n-1)}_f\Phi,B^{+(n-1)}_g\Phi\rangle\\
&=&4n(n-1)\bar{\rho}\sigma \Big[c\sum_{k=0}^{n-2}2^{2k+1}\frac{(n-1)!(n-2)!}{(((n-1)-k-1)!)^2}\langle f^{k+1},g^{k+1}\rangle\\
&&\langle B_f^{+((n-1)-k-1)}\Phi,B_g^{+((n-1)-k-1)}\Phi\rangle\Big]\\
&&+2nc\bar{\rho}\sigma|I|\langle B^{+(n-1)}_f\Phi,B^{+(n-1)}_g\Phi\rangle\\
&=&\Big[4n(n-1)\bar{\rho}\sigma +2nc\bar{\rho}\sigma|I|\Big]\langle B^{+(n-1)}_f\Phi,B^{+(n-1)}_g\Phi\rangle.
\end{eqnarray*}
This gives
\begin{eqnarray}\label{war}
\frac{\langle B^{+n}_f\Phi,B^{+n}_g\Phi\rangle}{(n!)^2}=4\bar{\rho}\sigma\Big[\frac{c|I|}{2n}+\frac{n-1}{n}\Big]\frac{\langle B^{+(n-1)}_f\Phi,B^{+(n-1)}_g\Phi\rangle}{((n-1)!)^2}.
\end{eqnarray}
On the other hand, if $I,\,J\subset\mathbb{R}^d$ such that $|I|<\infty,\,|J|<\infty$ and $I\cap J=\emptyset$, then by factorization property (see Theorem \ref{theo1}), one has 
\begin{eqnarray}\label{hed}
\langle\Psi_{\rho\chi_{I\cup J}},\Psi_{\sigma\chi_{I\cup J}}\rangle=\langle\Psi_{\rho\chi_I},\Psi_{\sigma\chi_I}\rangle\langle\Psi_{\rho\chi_J},\Psi_{\sigma\chi_ J}\rangle.
\end{eqnarray}
Therefore, if we put 
$$F(\rho,\sigma,|I|):=\langle\Psi_{\rho\chi_I},\Psi_{\sigma\chi_I}\rangle.$$
then from (\ref{war}), it follows that
$$F(\rho,\sigma,|I|+|J|)=\langle\Psi_{\rho\chi_{I\cup J}},\Psi_{\sigma\chi_{I\cup J}}\rangle.$$
for all $I,\,J\subset\mathbb{R}^d$ such that $|I|<\infty,\,|J|<\infty$ and $I\cap J=\emptyset$. Moreover, identity (\ref{hed}) implies that
$$F(\rho,\sigma,|I|+|J|)=F(\rho,\sigma,|I|)F(\rho,\sigma,|J|)$$
Thus, there must exists $\xi\in\mathbb{R}$ such that 
$$F(\rho,\sigma,|I|)=e^{|I|\xi}$$
Put $t=|I|$. The number $\xi$ is obtained by differentiating at $t=0$. 
Using relation (\ref{war}) one finds
$$
\frac{d}{dt}\Big|_{t=0}\frac{\langle B^{+n}_f\Phi,B^{+n}_g\Phi\rangle}{(n!)^2}
 = \frac{d}{dt}\Big|_{t=0}\Big(4\bar{\rho}\sigma(\frac{ct}{2n}+
\frac{n-1}{n})\dots4\bar{\rho}\sigma(\frac{ct}{2n}+0)\Big)
 = \frac{c}{2}\,\frac{(4\bar{\rho}\sigma)^n}{n}.
$$
Hence, one gets
$$
\xi = \frac{d}{dt}\Big|_{t=0}F(\rho,\sigma,t=|I|)
=\frac{c}{2}\sum_{n\geq1}\frac{(4\bar{\rho}\sigma)^n}{n}
 = -\frac{c}{2}\ln(1-4\bar{\rho}\sigma).
$$
This proves that the identity (\ref{identity}) holds true for any step functions\\
$f,\,g\in L^2(\mathbb{R}^d)\cap L^\infty(\Bbb R^d)$ such that $||f||_\infty<\frac{1}{2}$ 
and $||g||_\infty<\frac{1}{2}$.\\ 
By the factorization property (see Theorem \ref{theo2}), 
it is easy to show that, if $f=\sum_\alpha\rho_\alpha\chi_{I_{\alpha}}$,
$g=\sum_\beta\rho_\beta\chi_{I_{\beta}}\in L^2(\mathbb{R}^d)\cap L^\infty(\Bbb R^d)$, 
where $I_\alpha\cap I_{\alpha'}=\emptyset$ for all $\alpha\neq\alpha'$ 
and $|\rho_\alpha|<\frac{1}{2},\;|\rho_\beta|<\frac{1}{2}$ for all $\alpha,\,\beta$, 
then the equality  (\ref{identity}) is also satisfied.\\
Note that the set $\nu$ of functions $f=\sum_\alpha\rho_\alpha\chi_{I_{\alpha}}$ 
is a dense subset in\\
 $L^2(\mathbb{R}^d)\cap L^\infty(\Bbb R^d)$ equipped 
with the norm $\| \ \cdot \ \|=\| \ \cdot \ \|_\infty + \| \ \cdot \ \|_{2}$.\\
Consider now two functions $f$, $g$ in $L^2(\mathbb{R}^d)\cap L^\infty(\Bbb R^d)$ 
such that \\
$\|f\|_\infty, \|g\|_\infty<\frac{1}{2}$. 
Then, there exist $(f_j)_j,\;(g_j)_j\subset \nu$ and $j_0\in\mathbb{N}$ such that
$$
  \lim_{j\rightarrow\infty}||f_j-f||=0 \qquad ,\qquad 
  \lim_{j\rightarrow\infty}||g_j-g||=0\\
$$
$$
  ||f_j||_\infty<\frac{1}{2},\;||g_j||_\infty<\frac{1}{2}\qquad ; \ 
  \mbox{ for all }\,j\geq j_0.
$$
- {\bf{First step:}} Let us prove that, under the assumptions of the theorem, 
the integral
$$
e^{-\frac{c}{2}\int_{\mathbb{R}^d}\ln(1-4\bar{f}(s)g(s))ds}
$$
exists and
\begin{eqnarray}\label{Ro}
\lim_{j\rightarrow\infty}e^{-\frac{c}{2}\int_{\mathbb{R}^d}\ln(1-4\bar{f}_j(s)g_j(s))ds}=e^{-\frac{c}{2}\int_{\mathbb{R}^d}\ln(1-4\bar{f}(s)g(s))ds}.
\end{eqnarray}
For $z,z'\in\mathbb{C}$ such that $|z|<\frac{1}{4},\;|z'|<\frac{1}{4}$, we put
\begin{eqnarray*}
h(t)=\ln(1-4(tz+(1-t)z')),\;t\in[0,1].
\end{eqnarray*}
It is clear that $h$ is a derivable function on $[0,1]$. Hence, one has
$$h(1)-h(0)=\int_0^1h'(t)dt.$$
This gives
$$\ln(1-4z)-\ln(1-4z')=\Big(\int_0^1\frac{-4}{1-4(tz+(1-t)z')}dt\Big)(z-z').$$
It follows that
\begin{eqnarray}\label{Eq}
|\ln(1-4z)-\ln(1-4z')| &\leq & \sup_{t\in[0,1]}\Big|
\frac{4}{1-4(tz+(1-t)z')}\Big||z-z'|\nonumber\\
 &\leq &\frac{4}{1-4\sup_{t\in[0,1]}|tz+(1-t)z'|}|z-z'| \nonumber\\
 &\leq& \frac{4}{1-4\sup{(|z|,|z'|)}}|z-z'|.
\end{eqnarray} 

Note that if we take $z=\bar{f}(s)g(s)$ and $z'=0$, it is clear that $|z|<\frac{1}{4}$ and $|z'|<\frac{1}{4}$. Hence, identity (\ref{Eq}) implies that
\begin{eqnarray*}
|\ln(1-4\bar{f}(s)g(s))|&\leq&\frac{4}{1-4\sup\big(|\bar{f}(s)g(s)|,0\big)}|\bar{f}(s)g(s)|\\
&\leq&\frac{4}{1-4||f||_\infty||g||_\infty}|\bar{f}(s)g(s)|.
\end{eqnarray*}
This yields 
\begin{eqnarray*}
\Big|\int_{\mathbb{R}^d}\ln(1-4\bar{f}(s)g(s))ds\Big|&\leq&\frac{4}{1-4||f||_\infty||g||_\infty}\int_{\mathbb{R}^d}|\bar{f}(s)g(s)|ds\\
&\leq&\frac{4}{1-4||f||_\infty||g||_\infty}\|f\|_{2}\|g\|_{2}.
\end{eqnarray*}
Then, under the assumptions of the above theorem, the right hand side term of equality (\ref{identity}) exists.\\
Now, take $z=\bar{f}(s)g(s)$ and $z'=\bar{f}_j(s)g_j(s)_j$, for $j\geq j_0$. 
Then, it is easy to show that $|z|<\frac{1}{4}$ and $|z'|<\frac{1}{4}$. 
Moreover, from (\ref{Eq}), one has
\begin{eqnarray*}
&&\Big|\ln(1-4\bar{f}(s)g(s))-\ln(1-4\bar{f}_j(s)g_j(s))\Big|\\
&\leq&\frac{4}{1-4\sup{(|\bar{f}(s)g(s)|,|\bar{f}_j(s)g_j(s)|)}}\,\big|\bar{f}(s)g(s))-\bar{f}_j(s)g_j(s)\big|\\
&\leq&\frac{4}{1-4\sup{(||f||_\infty||g||_\infty,||f_j||_\infty||g_j||_\infty)}}\,\big|\bar{f}(s)g(s))-\bar{f}_j(s)g_j(s)\big|\\
&\leq&\frac{4}{1-4\sup{(||f||_\infty||g||_\infty,||f_j||_\infty||g_j||_\infty)}}\,\big|\bar{f}(s)g(s))-\bar{f}_j(s)g_j(s)\big|\\
&\leq&\frac{4}{1-4\sup{(||f||_\infty||g||_\infty,\sup_{j\geq j_0}(||f_j||_\infty||g_j||_\infty))}}\,\big|\bar{f}(s)g(s))-\bar{f}_j(s)g_j(s)\big|\\
&\leq&K\big|\bar{f}(s)g(s))-\bar{f}_j(s)g_j(s)\big|
\end{eqnarray*}
for all $j\geq j_0$, where $$K=\frac{4}{1-4\sup{(||f||_\infty||g||_\infty,\sup_{j\geq j_0}(||f_j||_\infty||g_j||_\infty))}}.$$
Therefore, for all $j\geq j_0$, one obtains
\begin{eqnarray*}
&&\big|\ln(1-4\bar{f}(s)g(s))-\ln(1-4\bar{f}_j(s)g_j(s))\big|\\
&\leq& K\big|\bar{f}(s)g(s))-\bar{f}(s)g_j(s))+\bar{f}(s)g_j(s))-\bar{f}_j(s)g_j(s)\big|\\
&\leq& K\Big(|\bar{f}(s)|\,|(g(s))-g_j(s)|+|\bar{f}(s)-\bar{f}_j(s)|\,|g_j(s)|\Big).
\end{eqnarray*}
This implies that 
\begin{eqnarray}\label{EE}
&&\Big|\int_{\mathbb{R}^d}\big(\ln(1-4\bar{f}(s)g(s))-\ln(1-4\bar{f}_j(s)g_j(s))\big)ds\Big|\nonumber\\
&&\leq K\int_{\mathbb{R}^d}|\bar{f}(s)|\,|g(s))-g_j(s|ds+K\int_{\mathbb{R}^d}|\bar{f}(s)-\bar{f}_j(s)|\,|g_j(s)|ds\nonumber\\
&&\leq K\Big(||f||_{2}||g-g_j||_{2}+||g_j||_{2}||f-f_j||_{2}\Big)
\end{eqnarray}
for all $j\geq j_0$. Note that it is clear that the term on the right hand side of (\ref{EE}) converges to 0, when $j$ tends to $\infty$. Thus, one gets
$$\lim_{j\rightarrow\infty}\int_{\mathbb{R}^d}\ln(1-4\bar{f}_j(s)g_j(s))ds=\int_{\mathbb{R}^d}\ln(1-4\bar{f}(s)g(s))ds.$$
This proves that
$$\lim_{j\rightarrow\infty}e^{-\frac{c}{2}\int_{\mathbb{R}^d}\ln(1-4\bar{f}_j(s)g_j(s))ds}=e^{-\frac{c}{2}\int_{\mathbb{R}^d}\ln(1-4\bar{f}(s)g(s))ds}.$$
- {\bf{Second step:}} The following identity holds. 
\begin{eqnarray}\label{Sam}
\lim_{j\rightarrow\infty}\langle \Psi(f_j),\Psi(g_j)\rangle=\langle \Psi(f),\Psi(g)\rangle.
\end{eqnarray}
In fact
$$\langle \Psi(f_j),\Psi(g_j)\rangle=\sum_{n\geq0}\frac{1}{(n!)^4}\langle B^{+n}_{f_j}\Phi,B^{+n}_{g_j}\Phi\rangle $$
for all $j\geq j_0$. Therefore, in order to prove (\ref{Sam}) 
it suffices to prove that
\begin{eqnarray}\label{ta}
\lim_{j\rightarrow\infty}\langle B^{+n}_{f_j}\Phi,B^{+n}_{g_j}\Phi\rangle=\langle B^{+n}_f\Phi,B^{+n}_g\Phi\rangle,
\end{eqnarray}
for all $n\in\mathbb{N}$. Let us prove (\ref{ta}) by induction.\\
- For $n=1$, one has
$$\langle B^+_{f_j}\Phi,B^+_{g_j}\Phi\rangle=c\langle f_j, g_j\rangle.$$
This implies that
\begin{eqnarray*}
\big|\langle B^+_{f_j}\Phi,B^+_{g_j}\Phi\rangle-\langle B^+_{f}\Phi,B^+_{g}\Phi\rangle\big| &=&c\big|\langle f_j, g_j\rangle-\langle f, g\rangle\big|\\
 &=&c\big|(\langle f_j, g_j\rangle-\langle f, g_j\rangle)+(\langle f, g_j\rangle-\langle f, g\rangle)\big|\\
 &\leq& c\|f-f_j\|_{2}\|g_j\|_{2}+c\|g-g_j\|_{2}\|f\|_{2}
\end{eqnarray*}  
which converges to $0$, when $j$ tends to $\infty$.\\
- Let $n\geq1$, suppose that
$$
\lim_{j\rightarrow\infty}\langle B^{+n}_{f_j}\Phi,B^{+n}_{g_j}\Phi\rangle=
\langle B^{+n}_{f}\Phi,B^{+n}_{g}\Phi\rangle.
$$
From Proposition \ref{prop1}, one has
\begin{eqnarray*}
&&\langle B^{+(n+1)}_{f_j}\Phi,B^{+(n+1)}_{g_j}\Phi\rangle-\langle B^{+(n+1)}_{f}\Phi,B^{+(n+1)}_{g}\Phi\rangle\\
&&=c\sum_{k=0}^n2^{2^k+1}\frac{(n+1)!n!}{(n-k)!}\\
&&\Big[\langle f_j^{k+1},g_j^{k+1}\rangle \langle B^{+(n-k)}_{f_j}\Phi,B^{+(n-k)}_{g_j}\Phi\rangle-\langle f^{k+1},g^{k+1}\rangle\langle B^{+(n-k)}_{f}\Phi,B^{+(n-k)}_{g}\Phi\rangle\Big]
\end{eqnarray*}
Note that
\begin{eqnarray}\label{Sa}
&&\Big|\langle f_j^{k+1},g_j^{k+1}\rangle \langle B^{+(n-k)}_{f_j}\Phi,B^{+(n-k)}_{g_j}\Phi\rangle-\langle f^{k+1},g^{k+1}\rangle\langle B^{+(n-k)}_{f}\Phi,B^{+(n-k)}_{g}\Phi\rangle\Big|\nonumber\\
&&=\Big|\langle f_j^{k+1},g_j^{k+1}\rangle \Big[\langle B^{+(n-k)}_{f_j}\Phi,B^{+(n-k)}_{g_j}\Phi\rangle-\langle B^{+(n-k)}_{f}\Phi,B^{+(n-k)}_{g}\Phi\rangle\Big]\nonumber\\
&&\;\;\;\;\;\;+\langle B^{+(n-k)}_{f}\Phi,B^{+(n-k)}_{g}\Phi\rangle\Big[\langle f_j^{k+1},g_j^{k+1}\rangle-\langle f^{k+1},g^{k+1}\rangle\Big]\Big|\nonumber\\
&&\leq|\langle f_j^{k+1},g_j^{k+1}\rangle|\Big|\langle B^{+(n-k)}_{f_j}\Phi,B^{+(n-k)}_{g_j}\Phi\rangle-\langle B^{+(n-k)}_{f}\Phi,B^{+(n-k)}_{g}\Phi\rangle\Big|\nonumber\\
&&\;\;\;\;\;\;+\big|\langle B^{+(n-k)}_{f}\Phi,B^{+(n-k)}_{g}\Phi\rangle
\big|\Big|\langle f_j^{k+1},g_j^{k+1}\rangle-\langle f^{k+1},g^{k+1}\rangle\Big|
\end{eqnarray}
and from the induction assumption it follows that
\begin{eqnarray}\label{Am}
\lim_{j\rightarrow\infty}\langle B^{+(n-k)}_{f_j}\Phi,B^{+(n-k)}_{g_j}\Phi\rangle
=\langle B^{+(n-k)}_{f}\Phi,B^{+(n-k)}_{g}\Phi\rangle
\end{eqnarray}
for all $k=0,1,...,n$. Now, let us prove that
\begin{eqnarray}\label{Rg}
\lim_{j\rightarrow\infty}\langle f_j^{k+1},g_j^{k+1}\rangle=\langle f^{k+1},g^{k+1}\rangle.
\end{eqnarray}
One has
\begin{eqnarray*}
\langle f_j^{k+1},g_j^{k+1}\rangle-\langle f^{k+1},g^{k+1}\rangle=\langle f_j^{k+1}-f^{k+1},g_j^{k+1}\rangle+\langle f^{k+1},g_j^{k+1}-g^{k+1}\rangle.
\end{eqnarray*} 
This gives
$$
\Big|\langle f_j^{k+1},g_j^{k+1}\rangle-\langle f^{k+1},g^{k+1}\rangle\Big| \leq 
\|g_j^{k+1}\|_{2}\|f_j^{k+1}-f^{k+1}\|_{2} +\|f^{k+1}\|_{2}\|g_j^{k+1}-g^{k+1}\|_{2}
.$$
Notice that
$$f_j^{k+1}-f^{k+1}=(f_j-f)v_j.$$
where $v_j\in L^\infty(\mathbb{R}^d)$ is a function on $\mathbb{R}^d$, which depends on $f$ and $f_j$ such that
$$\sup_j\|v_j\|_\infty<\infty$$
It follows that
$$\|f_j^{k+1}-f^{k+1}\|_{2}\leq\sup_j\|v_j\|_\infty\|f_j-f\|_{2} $$
which converges to 0, when $j$ goes to $\infty$. Thus, the relation (\ref{Rg}) holds true. Using all together (\ref{Sa}), (\ref{Am}) and (\ref{Rg}), we have proved that
$$\lim_{j\rightarrow0}\langle B^{+n}_{f_j}\Phi,B^{+n}_{g_j}\Phi\rangle=\langle B^{+n}_{f}\Phi,B^{+n}_{g}\Phi\rangle $$
for all $n\in\mathbb{N}$. This implies that there exists $j_1\in\mathbb{N}$ such that for all $n\in\mathbb{N}$
$$\Big|\langle B^{+n}_{f_j}\Phi,B^{+n}_{g_j}\Phi\rangle-\langle B^{+n}_{f}\Phi,B^{+n}_{g}\Phi\rangle\Big|\leq\frac{1}{j} $$
$\forall j\geq j_1$.
Hence, for all $j\geq \max{(j_0,j_1)}$, one obtains
\begin{eqnarray*}
\big|\langle \Psi(f_j),\Psi(g_j)\rangle-\langle \Psi(f),\Psi(g)\rangle\big|&\leq&\sum_{n\geq0}\frac{1}{(n!)^4}\big|\langle B^{+n}_{f_j}\Phi,B^{+n}_{g_j}\Phi\rangle-\langle B^{+n}_{f}\Phi,B^{+n}_{g}\Phi\rangle\big|\\
&\leq&\frac{1}{j}\sum_{n\geq0}\frac{1}{(n!)^4}
\end{eqnarray*}
which converges to 0, when $j$ tends to $\infty$.\\
In conclusion, from (\ref{Ro}) and (\ref{Sam}), one deduces that 
\begin{eqnarray*}
\lim_{j\rightarrow\infty}e^{-\frac{c}{2}\int_{\mathbb{R}^d}\ln(1-4\bar{f}_j(s)g_j(s))ds}&=&e^{-\frac{c}{2}\int_{\mathbb{R}^d}\ln(1-4\bar{f}(s)g(s))ds},\\
\lim_{j\rightarrow\infty}\langle \Psi(f_j),\Psi(g_j)\rangle&=&\langle \Psi(f),\Psi(g)\rangle.
\end{eqnarray*}
But, for all $j\geq j_0$
$$
\langle\Psi(f_j),\Psi(g_j)\rangle=
e^{-\frac{c}{2}\int_{\mathbb{R}^d}\ln(1-4\bar{f}_j(s)g_j(s))ds}
.$$
This implies that
\begin{eqnarray*}
\langle \Psi(f),\Psi(g)\rangle=e^{-\frac{c}{2}\int_{\mathbb{R}^d}\ln(1-4\bar{f}(s)g(s))ds}.
\end{eqnarray*}
\end{proof}

\section{Linear independence of the quadratic exponential vectors}
\label{Lin-ind-q-exp-vect}

The following lemma is an immediate consequence of the Schur Lemma.
\begin{lemma}\label{boukas}
If $A=(a_{i,j})_{1\leq i,j\leq N}$ be a positive matrix. 
Then, for all $n\in \mathbb{N}$, the matrix $((a_{i,j})^n)_{1\leq i,j\leq N}$ 
is also positive. In particular, $(e^{a_{i,j}})_{1\leq i,j\leq N}$ is a positive matrix.
\end{lemma}
Now, we prove the following.
\begin{lemma}\label{rolando}
Let $f_1,\,\dots,f_N$ be functions in $L^2(\mathbb{R}^d)\cap L^\infty(\mathbb{R}^d)$. Suppose that for all $i,j=1,\dots,N$, there exists a subset $I_{i,j}$ of $\mathbb{R}^d$ such that $|I_{i,j}|>0$ and $f_i(x)\neq f_j(x)$ for all $x\in I_{i,j}$. Suppose that $|supp(f_i)|>0$, for all $i=1,\dots,N$. Then, the identity
\begin{eqnarray}\label{barc}
\lambda_1(f_1(x))^n+\dots+\lambda_N(f_N(x))^n=0
\end{eqnarray}
for all $n\in\mathbb{N}$ and for almost any $x\in\mathbb{R}^d$, implies that $\lambda_1=\dots=\lambda_N=0.$
\end{lemma} 
\begin{proof} 
By induction. If $N=2$, then one has 
\begin{eqnarray}\label{Boukas}
\lambda_1(f_1(x))^n+\lambda_2(f_2(x))^n=0\qquad ;\,a.e \;x\in\mathbb{R}^d
\end{eqnarray}
Suppose that $\lambda_1\neq0$. We can assume that there exists $x_0\in\mathbb{R}^d$ 
such that $|f_1(x_0)|>|f_2(x_0)|$. From (\ref{Boukas}), it is clear that $\lambda_2\neq0$ 
and $f_2(x_0)\neq0$. Hence, one gets
$$
\frac{\lambda_1}{\lambda_2}=
-\lim_{n\rightarrow\infty}\Big(\frac{f_2(x_0)}{f_1(x_0)}\Big)^n=0
$$
This yields that $\lambda_1=0$, which is impossible by assumption.\\
Let $N\geq2$. Let $f_1,\,\dots,f_N$ in $L^2(\mathbb{R}^d)\cap L^\infty(\mathbb{R}^d)$, 
which satisfy the hypotheses of the lemma. 
Suppose that if identity (\ref{barc}) holds, then $\lambda_1=\dots=\lambda_N=0.$ \\
Now, consider $f_1,\,\dots,f_N,f_{N+1}$ in $L^2(\mathbb{R}^d)\cap L^\infty(\mathbb{R}^d)$,
which satisfy the hypotheses of the lemma and assume that 
\begin{eqnarray}\label{nab}
\lambda_1(f_1(x))^n+\dots+\lambda_N(f_N(x))^n+\lambda_{N+1}(f_{N+1}(x))^n=0
\end{eqnarray}
for all $n\in\mathbb{N}$ and for almost any $x\in\mathbb{R}^d$. 
Because the hypotheses satisfied by $f_1,\,\dots,f_N,f_{N+1}$, there exists $x_0\in\mathbb{R}^d$ such that for some $i_0\in\{1,\dots,N+1\}$, one has 
$$
|f_{i_0}(x)|>|f_i(x_0)| 
$$
for all $i\in\{1,\dots,N+1\}$ and $i\neq i_0$. Without loss of generality suppose 
that $i_0=N+1$. So, identity (\ref{nab}) implies that
$$
\lambda_{N+1}=-\lim_{n\rightarrow\infty}\Big[\lambda_1\Big(\frac{f_1(x_0)}{f_{N+1}(x_0)}\Big)^n+\dots+\lambda_N\Big(\frac{f_N(x_0)}{f_{N+1}(x_0)}\Big)^n\Big]=0.
$$
Hence, one gets $\lambda_{N+1}=0$ and, by the induction assumption, 
one can conclude that $\lambda_1=\dots=\lambda_N=0.$ 
\end{proof}

As a consequence of Lemmas \ref{boukas}, \ref{rolando}, 
we prove the following theorem.
\begin{theorem}\label{LI-quad-exp-v}
Let $f_1,\,\dots,f_N$ be functions in $L^2(\mathbb{R}^d)\cap L^\infty(\mathbb{R}^d)$ such that $\|f_i\|_\infty<\frac{1}{2}$ for all $i=1,\dots,N$. Suppose that for all $i,j=1,\dots,N$, there exists a subset $I_{i,j}$ of $\mathbb{R}^d$ such that $|I_{i,j}|>0$ and $f_i(x)\neq f_j(x)$ for all $x\in I_{i,j}$. Then, the quadratic exponential vectors $\Psi(f_1),\dots,\Psi(f_N)$ are linearly independents.
\end{theorem}
\begin{proof}
Let $A=(a_{ij})_{i,j=1}^N,$ where
$$a_{i,j}=-\int_{\mathbb{R}^d}\ln(1-\bar{f}_i(x)f_j(x))dx.$$ 
Then, one has
\begin{eqnarray}\label{rodrigo}
\sum_{i,j=1}^N\bar{\lambda}_i\lambda_ja_{i,j}&=&-\sum_{i,j=1}^N\bar{\lambda}_i\lambda_j\int_{\mathbb{R}^d}\ln(1-\bar{f}_i(x)f_j(x))dx\nonumber\\
&=&\int_{\mathbb{R}^d}\sum_{n\geq1}\frac{1}{n}\Big[\sum_{i,j=1}^N\bar{\lambda}_i\lambda_j(\bar{f}_i(x))^n(f_j(x)))^n\Big]dx\nonumber\\
&=&\!\!\int_{\mathbb{R}^d}\sum_{n\geq1}\frac{1}{n}\Big|\lambda_1(f_1(x))^n+\dots+\lambda_N(f_N(x))^n\Big|^2dx\geq0.
\end{eqnarray}
Thus $A$ is positive definite. 
Now, let $\lambda_1,\dots,\lambda_N$ be scalars such that
\begin{equation}\label{slm}
\lambda_1\Psi(f_1)+\dots+\lambda_n\Psi(f_N)=0.
\end{equation}
Then, identity (\ref{slm}) holds if and only if 
$$\|\lambda_1\Psi(f_1)+\dots+\lambda_N\Psi(f_N)\|^2=0.$$ 
That is
\begin{eqnarray}\label{ro}
\sum_{i,j=1}^N\bar{\lambda}_i\lambda_j\langle \Psi(f_i),\Psi(f_j)\rangle=\sum_{i,j=1}^N\bar{\lambda}_i\lambda_jb_{i,j}=0,
\end{eqnarray}
where
$$b_{i,j}=e^{a_{i,j}}=e^{-\int_{\mathbb{R}^d}\ln(1-\bar{f}_i(x)f_j(x))dx}=\langle \Psi(f_i),\Psi(f_j)\rangle.$$
Note that idendity (\ref{ro}) implies that
\begin{eqnarray}\label{Ahm}
\sum_{n\geq0}\frac{1}{n!}\Big[\sum_{i,j=1}^N\bar{\lambda}_i\lambda_j(a_{i,j})^n\Big]=0.
\end{eqnarray}
Recall that $A$ is a positive matrix. Then, Lemma \ref{boukas} implies that for all $n\in\mathbb{N}$, the matrix $((a_{i,j})^n)_{1\leq i,j\leq N}$ is also positive. So, from (\ref{Ahm}), one has 
$$\sum_{i,j=1}^N\bar{\lambda}_i\lambda_j(a_{i,j})^n=0 $$
for all $n\in\mathbb{N}$. In particular,
$$\sum_{i,j=1}^N\bar{\lambda}_i\lambda_ja_{i,j}=0.$$
Therefore, from (\ref{rodrigo}), one gets
\begin{eqnarray}\label{zoh}
\lambda_1(f_1(x))^n+\dots+\lambda_N(f_N(x))^n=0,
\end{eqnarray}
for all $n\in\mathbb{N}$ and for almost everywhere $x\in\mathbb{R}^d$.

- First case: For all $i,j=1,\dots,N$, there exists a subset $I_{i,j}$ of $\mathbb{R}^d$ such that $|I_{i,j}|>0$, $f_i(x)\neq f_j(x)$ for all $x\in I_{i,j}$ and $|supp(f_i)|>0$, for all $i=1,\dots,N$. Then, identity (\ref{zoh}) and Lemma \ref{rolando} imply that $\lambda_1=\dots\lambda_N=0.$ 

- Second case: There exists $i_0\in\{1,\dots,N\}$ such that $f_{i_0}=0$, a.e. From the assumptions of the above theorem, it is clear that $|supp(f_i)|>0$ for all $i\neq i_0$. Without loss of generality, suppose that $i_0=N$. Then, identity (\ref{zoh}) becomes
$$
\lambda_1(f_1(x))^n+\dots+\lambda_{N-1}(f_{N-1}(x))^n=0
$$
for all $n\in\mathbb{N}$ and for almost everywhere $x\in\mathbb{R}^d$. Moreover, the functions $f_1,\dots,f_{N-1}$ satisfy the hypotheses of Lemma \ref{rolando}, which implies that $\lambda_1=\dots\lambda_{N-1}=0.$ Now, taking account of (\ref{slm}), one obtains that $\lambda_1=\dots\lambda_N=0.$ This ends the proof.
\end{proof}
\section{Property of the family set of quadratic exponential vectors}
In this section, we prove that the set of the quadratic exponential vectors 
is a total set in the quadratic Fock space.
\begin{theorem}
The set of the quadratic exponential vectors is a total set in the quadratic Fock space.
 Moreover, one has
\begin{equation}\label{ism}
B^{+n}_f\Phi=\frac{d^n}{dt^n}\Big|_{t=0}\Psi(tf)
\end{equation}
for all $f\in L^2(\mathbb{R}^d)\cap L^\infty(\mathbb{R}^d)$.
\end{theorem}
\begin{proof}
If $\|f\|_\infty=0$ then (\ref{ism}) is  clearly verified. 
Therefore we can assume that $f\in L^2(\mathbb{R}^d)\cap L^\infty(\mathbb{R}^d)$ 
such that $\|f\|_\infty>0$. Consider $0\leq t<\delta$ with 
$$
\delta<\frac{1}{2\|f\|_\infty}.
$$
Recall that for all $0\leq t<\delta$, one has
$$
\Psi(tf)=\sum_{m\geq0}\frac{t^m}{m!}B^{+n}_f\Phi.
$$
Then, for all $m>n$, one has
$$
\frac{d^n}{dt^n}\Big(\frac{t^m}{m!}B^{+m}_f\Phi\Big)=\frac{t^{m-n}}{(m-n)!}B^{+m}_f\Phi
$$
It is obvious that for all $0\leq t<\delta$ the quadratic exponential exists and one has
$$
\|\frac{t^{m-n}}{(m-n)!}B^{+m}_f\Phi\|\leq \frac{\delta^{m-n}}{(m-n)!}\|B^{+m}_f\Phi\|
=: U_m.
$$
But, from identity (\ref{you}), one has
\begin{eqnarray*}
||B^{+m}_f\Phi||^2\leq\Big[4m(m-1)\|f\|^2_\infty+2m\|f\|^2_2\Big]\|B^{+(m-1)}_f\Phi\|^2.
\end{eqnarray*}
This proves 
$$
U_m\leq\frac{\sqrt{4m(m-1)\|f\|^2_\infty+2m\|f\|^2_2}}{m-n}\,\delta \;U_{m-1}.
$$
This yields 
$$
\lim_{m\rightarrow\infty}\frac{U_m}{U_{m-1}}\leq 2\|f\|_\infty\delta<1.
$$
Then, the series $\sum_mU_m$ converges. Therefore, one gets
$$
\frac{d^n}{dt^n}\Big(\sum_{m\geq0}\frac{t^m}{m!}B^{+n}_f\Phi\Big)
=\sum_{m\geq n}\frac{d^n}{dt^n}\Big(\frac{t^m}{m!}B^{+m}_f\Phi\Big)
=\sum_{m\geq n}\frac{t^{m-n}}{(m-n)!}B^{+m}_f\Phi.
$$
Finally, by taking $t=0$ the result of the above theorem holds.
\end{proof}


\begin{thebibliography}{99}
\bibitem[1]{ALV}L. Accardi, Y. G. Lu and I. V. Volovich: White noise approach to classical and quantum stochastic calculi, {\it Centro Vito Volterra, Universit\`a di Roma ``Tor Vergata'', preprint 375, 1999.}
\bibitem[2]{ADS}L. Accardi, A. Dhahri and M. Skeide: Extension of quadratic exponential vectors, submitted to: Proceedings of the  {\it 29-th Conference on Quantum Probability and related topics }, Hammamet (2008).
\bibitem[2b]{AS}L. Accardi, M. Skeide: On the relation of the Square of White Noise and the Finite Difference Algebra,
IDA--QP (Infinite Dimensional Analysis, Quantum Probability and Related Topics) 3 (2000) 185--189
Volterra Preprint  N. 386 (1999)
\bibitem[3]{AcDha09a} L. Accardi, A. Dhahri: The quadratic Fock functor, {\it 
submitted}.
\bibitem[4]{AAF}L. Accardi, G. Amosov and U. Franz: Second quantization automorphisms of the renormalized square of white noise (RSWN) algebra, {\it Inf. Dim. Anl, Quantum Probability and related topics, Vol. 7, No. 2 (2004) 183-194}.
\bibitem[5]{AFS}L. Accardi, U. Franz and M. Skeide: Renormalized squares of white noise and other non-Gaussian noises as Levy processes on real Lie algebras, {\it Commun. Math. Phys. {\bf 228} (2002) 123-150}.
\bibitem[6]{P} K. R. Parthasarathy: {{\it An Introduction to Quantum Stochastic Calculus}. Birkh\"auser Verlag: Basel. Boston. Berlin.}
\bibitem[7]{[AcKuSt09sigma]} L. Accardi, A. Boukas
Quantum probability, renormalization and infinite dimensional $*$--Lie algebras, 
SIGMA (Symmetry, Integrability and Geometry: Methods and Applications) (2009),
electronic journal, Special issue on: Kac-Moody Algebras and Applications 
$http://www.emis.de/journals/SIGMA/Kac-Moody-algebras.html$

\end{thebibliography}
\end{document}